\documentclass{revtex4}

\usepackage{graphicx}
\usepackage{dcolumn}
\usepackage{amsmath}

\begin{document}
\bibliographystyle{revtex}


\title[Short Title]{Dynamical aspects of isoscaling}

\author{C. O. Dorso${}^1$, C. R. Escudero${}^2$, M. Ison${}^1$, and J. A. L\'opez${}^2$}

\affiliation{${}^1$Departamento de F\'isica, FCEN, Universidad de Buenos Aires, N\'u\~nez, Argentina \\
${}^2$Department of Physics, University of Texas at El Paso, El
Paso, Texas 79968, U.S.A.}

\date{\today}

\begin{abstract}
The origin and dynamical evolution of isoscaling was studied using
classical molecular dynamics simulations of ${}^{40}$Ca +
${}^{40}$Ca, ${}^{48}$Ca + ${}^{48}$Ca, and ${}^{52}$Ca +
${}^{52}$Ca, at beam energies ranging from $20 \ MeV/A$ to $85 \
MeV/A$. The analysis included a study of the time evolution of this
effect. Isoscaling was observed to exist in these reactions from the
very early primary isotope distributions (produced by highly {\it
non-equilibrated} systems) all the way to asymptotic times. This
indicates that isoscaling is independent of quantum effects and
thermodynamical equilibrium. In summary, collision-produced
isoscaling appears to be due more to the mere partitioning of the
proton-neutron content of the participant nuclei, than to specific
details of the reaction dynamics.

\end{abstract}

\maketitle

\section{Introduction}\label{intro}

Experimental advances, that now permit the study of nuclear
reactions involving radioactive isotopes, have propelled the
isotopic degree of freedom to the
forefront~\cite{xu,johnston,laforest,tsang,tsang2001}.  It is
expected, for instance, that this new variable of control could
shed light on charge equilibration in heavy ion reactions, and on
the role played by the isotope asymmetry terms of the equation of
state of nuclear matter~\cite{tsang}.

The main tool for inspection of this new observable is based on
the isotope yields of central collisions of similar, but
isotopically different, reactions~\cite{xu,tsang2001}.  The ratio
of isotope yields from reactions $1$ and $2$, $R_{21}(N, Z)$, has
been found to depend exponentially on the isotope neutron number
N, and proton number, Z:
\begin{equation}\label{r12}
R_{21}(N,Z)=  {\frac{Y_2(N, Z)}{Y_1(N, Z)}} \approx e^{\alpha N +
\beta Z} \ ,
\end{equation}
where $\alpha$ and $\beta$ are fitting parameters. Equations of
the form of (\ref{r12}) can be linked, under some approximations,
to primary isotope yields produced by disassembling infinite
equilibrated systems in microcanonical and grand canonical
ensembles~\cite{tsang}, as well as to breakups in
canonical~\cite{das} ensembles.

Little reflection is needed to understand that $R_{21}$ could be
affected by many reaction variables.  Its direct dependence on the
experimentally measured yields, $Y_1$ and $Y_2$, makes $R_{21}$
vulnerable to anything that can modify the isotopic content of the
initial fragment yield, be this out-of-equilibrium breakup,
secondary fission of primordial fragments, light particle
emission, $\beta$-decay, and the like. As these effects have
varying lifetimes, it is very likely that the
experimentally-captured yield contains an integral of all of these
effects on the primary distribution.  This casts a shadow of doubt
on the isoscaling-related conclusions obtained by the use of
microcanonical, canonical and grand canonical breakup scenarios,
and calls for the use of an unconstrained model to ratify the main
findings of equilibrium models.

This article aims at elucidating the origin and dynamical evolution
of isoscaling using a model not restricted by assumptions such as
the existence of freeze-out stages, thermal and chemical
equilibration, or unrealistic volume constraints~\cite{furuta}. In
the next section, the $MD$ model used is introduced along with the
fragment recognition algorithm selected for this study.  The fitting
procedure used to extract the isoscaling exponential law~(\ref{r12})
from the simulations is presented in section \ref{isoscaling}
followed by results consisting of the observation of isoscaling at
asymptotic times, its dependence on the reaction energy, the time
evolution of such a law during the reaction, and its connection to
thermodynamical observables. The manuscript closes with a summary of
the main conclusions.

\section{Molecular dynamics}\label{moldyn}
To study the origin of isoscaling, a model capable of reproducing
both the out-of-equilibrium and the equilibrium parts of a collision
is needed.  As statistical and other equilibrium
models~\cite{smm,ees} lack -by construction- of all relevant
collision-induced correlations, a dynamical model is thus needed.
Most of such dynamical models, nevertheless, either lack
higher-order correlations and have difficulties in producing
fragmentation~\cite{beg88,dan,bao,aij86}, or introduce unrealistic
volume constraints to the dynamics~\cite{ono}. In the present work,
we use a molecular dynamics ($MD$) model that can describe
non-equilibrium dynamics, hydrodynamic flow and changes of phase
without adjustable parameters.  The combination of this $MD$ code
with a fragment-recognition algorithm, has been dubbed
``$Latino$''~\cite{latino}, and in recent years it has been applied
successfully to study, among other things, neck
fragmentation~\cite{chernolaval}, phase transitions~\cite{oax2001},
critical phenomena~\cite{HIP2003,CIT} and the caloric
curve~\cite{barra2004,brasil} in nuclear reactions.

The $MD$ code uses a two-body potential composed of the Coulomb
interaction plus a nuclear part~\cite{pandha} that correctly
reproduces nucleon-nucleon cross sections, as well as the correct
binding energies and densities of real nuclei. The ``nuclear''
part of the interaction potential is
\begin{eqnarray}
V_{np}(r) &=&V_r\left[ exp(-\mu _rr)/{r} -
exp(-\mu _rr_c)/{r_c}\right] \nonumber \\
&&\ \mbox{}-V_a\left[ exp(-\mu _ar)/{r} -
exp(-\mu _ar_a)/{r_a}\right] \nonumber \\
V_{nn}(r) &=&V_{pp}(r)=V_0\left[ exp(-\mu _0r)/{r} - exp(-\mu
_0r_c)/{r_c}
 \right]
\label{2BP}
\end{eqnarray}
where the cutoff radius is $r_c=5.4~fm$, $V_{np}$ is the potential
between a neutron and a proton while $V_{nn}$ is that between
identical nucleons. The values of the parameters of the Yukawa
potentials~\cite{pandha} correspond to an equation of state of
infinite nuclear matter with an equilibrium density of $\rho
_0=0.16~fm^{-3}$, a binding energy $E(\rho _0)=-16~MeV/nucleon$,
and a compressibility of around $250~MeV$.

To study collisions, this potential is first used along with
dissipative molecular dynamics to construct ``nuclei'' by grouping
``nucleons'' at the binding energies and radii of real nuclei, and
stable for times longer than the reaction time.  These nuclei are
then used as targets and projectiles by rotating the relative
orientation of the target-nuclei combination, boosting the
center-of-mass velocity of the projectile to a desired energy, and
leaving the target initially at rest.  The trajectories of motion of
individual nucleons are then calculated using the standard Verlet
algorithm with an energy conservation of $\mathcal{O}$($0.01 \%)$.

The collision information, initially composed by the values of
$(\overrightarrow{r},\overrightarrow{p})$ of the nucleons, is then
transformed into fragment information by means of a
cluster-detection algorithm which, in this case, is the $MSE$ method
which was introduced decades ago~\cite{hill}, but was recently
adapted for this field, fully analyzed and compared with other
fragment recognition algorithms~\cite{mse}. According to this
prescription, a particle $i$ belongs to a cluster $C$ if there is a
particle $j$ in $C$ to which $i$ is bound in the sense of
$p_{ij}^{2}/{4\mu}<v_{ij}$, where $p_{ij}$ is the relative momentum,
$\mu$ the reduced mass, and $v_{ij}$ the interparticle potential. In
this cluster definition the effect of the relative momentum between
the particles that form the cluster is taken into account in an
approximate way; of course, at some time during the reaction, $MSE$
yields the asymptotic cluster distribution. The resulting
information contains details about the nucleon content of the
emitted fragments and, as this is available at all times during the
collision, it allows the study of the time evolution of quantities
such as the isoscaling law.

Before we turn to a description of the analysis performed on these
collisions, a word of caution is needed to underline the fact that
the $MD$ model here described is fully classical.  Since all
quantal effects (such as the exclusion principle) and isotopic
content-modifying phenomena (such as $\beta$-decay) are excluded,
means that, if any of them is responsible for isoscaling, this
study should not reproduce this effect.  On the contrary, if
isoscaling is well predicted by this classical model, this will
exclude any purely quantum phenomenon as the sole cause of
isotopic scaling.

\begin{figure}[htb]
\par
\centerline{ \includegraphics[scale=0.5,angle=-90]{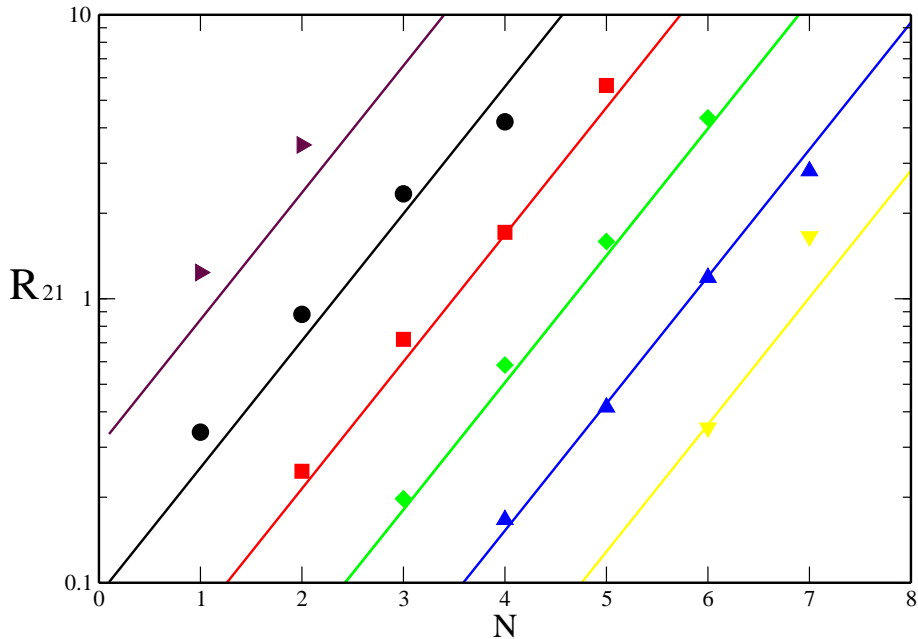}}
\caption{\label{fig1}Asymptotic behavior of isoscaling.  Typical fit
to $R_{21}(N,Z)$ for the case ${}^{40}$Ca - ${}^{48}$Ca at $35 \
MeV/A$ at $1250 \ fm/c$.}
\end{figure}

\section{Isoscaling}\label{isoscaling}
The collisions ${}^{40}$Ca+${}^{40}$Ca, ${}^{48}$Ca+${}^{48}$Ca, and
${}^{52}$Ca+${}^{52}$Ca, were studied at beam energies of $20$,
$25$, $35$, $45$, $65$ and $85 \ MeV/A$ with two thousand collisions
performed at each energy.  Data from these collisions were used to
construct the corresponding yield matrices $Y_i(N,Z)$, where $i$
stands for the reaction, and $N$ and $Z$ for the neutron and proton
numbers, respectively. These matrices were then used to calculate
the ratio $R_{21}(N,Z)=Y_2(N,Z)/Y_1(N,Z)$ for the combinations of
reactions ${}^{40}$Ca with ${}^{48}$Ca, ${}^{40}$Ca with
${}^{52}$Ca, and ${}^{48}$Ca with ${}^{52}$Ca at each of the energy
values.  These ratios were calculated at different reaction times
starting at impact ($t=0$) and ending at $5000 \ fm/c$; this last
time corresponds, practically, to an asymptotic value. Fits to the
isoscaling exponential law (\ref{r12}) were obtained using a
standard least squares method for each reaction and energy,
procedure which yielded values of the parameters $\alpha$ and
$\beta$ for each of the calculated times.  Next, studies of the
energy dependence of isoscaling at long times and its behavior
during the dense part of the reaction are presented.

\subsection{Isoscaling in asymptotia}\label{asymp}
Experimental results always correspond to asymptotic values,
taking into account that, at the energies listed below, the
crossing time is of the order of a few {\it fm/c}, it is safe to
consider times of, say, $1000 \ fm/c$ as asymptotic; in this
study, nevertheless, the calculation was extended to five times
this value.

\begin{figure}[htb]
\par
 \centerline{ \includegraphics[scale=0.5,angle=-90]{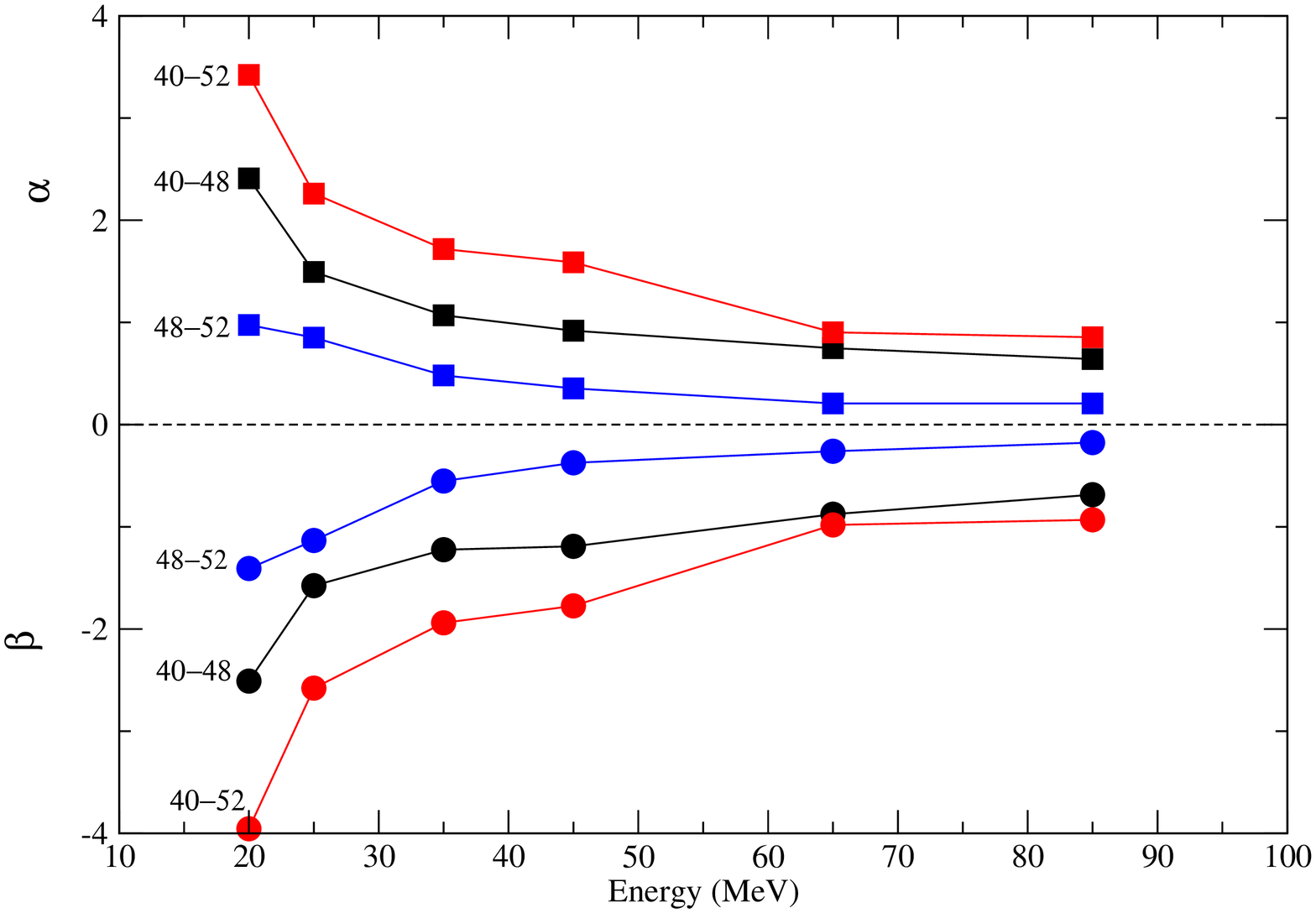}}
 \caption{\label{fig2}Energy dependence of
fitting parameters $\alpha$ and $\beta$ for three reactions
${}^{40}$Ca+${}^{40}$Ca, ${}^{48}$Ca+${}^{48}$Ca, and
${}^{52}$Ca+${}^{52}$Ca, at beam energies ranging from $20$ to $85
\ MeV/A$ and at $t=5000 \ fm/c$.}
\end{figure}

Figure~\ref{fig1} shows the obtained values of $R_{21}(N,Z)$ and the
corresponding fit to the isoscaling exponential law (\ref{r12}) for
the case ${}^{40}$Ca - ${}^{48}$Ca at $35 \ MeV/A$ at $1250 \ fm/c$.
The fact that $Latino$ reproduces the reactions sufficiently well as
to produce isoscaling at long times is obvious from this figure.
Other reactions ({\it i.e.} with other masses and energies) yield
similar results.

Upon applying the fitting procedures described in~\ref{isoscaling}
for each of the three ratios of reactions, values of $\alpha$ and
$\beta$ were obtained for each of the energies. Figure~\ref{fig2}
shows the values of these fitting parameters at asymptotic times
($5000 \ fm/c$) as a function of beam energy per nucleon for the
ratios of the reactions ${}^{40}$Ca - ${}^{48}$Ca, ${}^{40}$Ca -
${}^{52}$Ca, and ${}^{48}$Ca - ${}^{52}$Ca.  The smooth trends of
$\alpha$ and $\beta$ are in the range of values obtained by other
investigators~\cite{tsang,tsang2001,tsang2002,ono,tsang111}.

It is instructive to compare these results to those recently
obtained using quantal molecular dynamics simulations with the
$AMD$-$V$ model~\cite{ono}. In that work, collisions between
${}^{40}$Ca$+^{40}$Ca, ${}^{48}$Ca$+^{48}$Ca and
${}^{60}$Ca$+^{60}$Ca at $35 \ MeV/A$ were performed, and an
isoscaling-looking behavior was observed for their $R_{21}$. The
power-law fit obtained for the ratio of $^{48}Ca+^{48}Ca$ to of
$^{40}Ca+^{40}Ca$ yielded an $\alpha=1.03$ and $\beta =$ 1.22. These
values agree with the results here obtained here for a similar case:
$\alpha =1.07$ and $\beta= 1.22$ ({\it cf.} figure~\ref{fig2});
other $AMD$-$V$ results are also in line with our classical $MD$
simulations.

As stated before, since this study is based on a classical $MD$
model, the fact that isospin is well reproduced by it, implies
that this effect is not due to quantum effects that could modify
the isotopic content of the fragments, such as $\beta$-decay. To
gain a deeper insight into the origin of isoscaling, we now turn
to a study to its time evolution during the reaction.

\begin{figure}[htb]
\par
\centerline{ \includegraphics[scale=0.5,angle=-90]{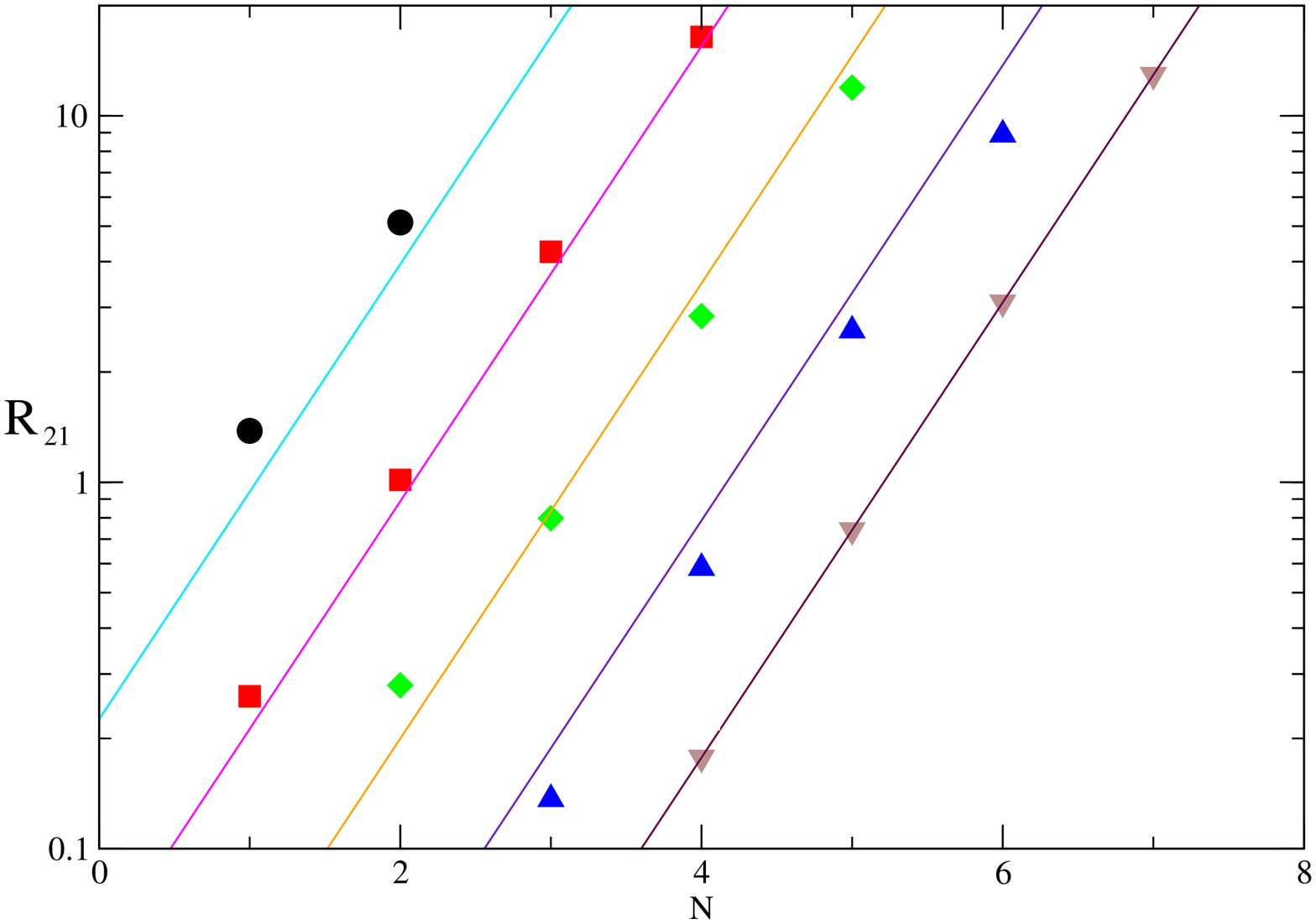}}
 \caption{\label{fig3}Early behavior of isoscaling.  Typical fit to $R_{21}(N,Z)$ for
the case ${}^{40}$Ca - ${}^{48}$Ca at $35 \ MeV/A$ at $125 \ fm/c$.}
\end{figure}

\subsection{Evolution of isoscaling}\label{evolution}
As the nucleon information is available throughout the reaction, the
yields and, thus, the ratio $R_{21}$ can be calculated at any time
during the collision. Figure~\ref{fig3} shows early (at $125 \
fm/c$) values of $R_{21}(N,Z)$ and the corresponding fit to the
isoscaling exponential law (\ref{r12}), for the case ${}^{40}$Ca -
${}^{48}$Ca at $35 \ MeV/A$. The observed goodness of the isoscaling
exponential law appears to be as good in early times as in
asymptotic times. Other reactions yield similar results.

\begin{figure}[htb]
\par
\centerline{ \includegraphics[scale=0.5,angle=-90]{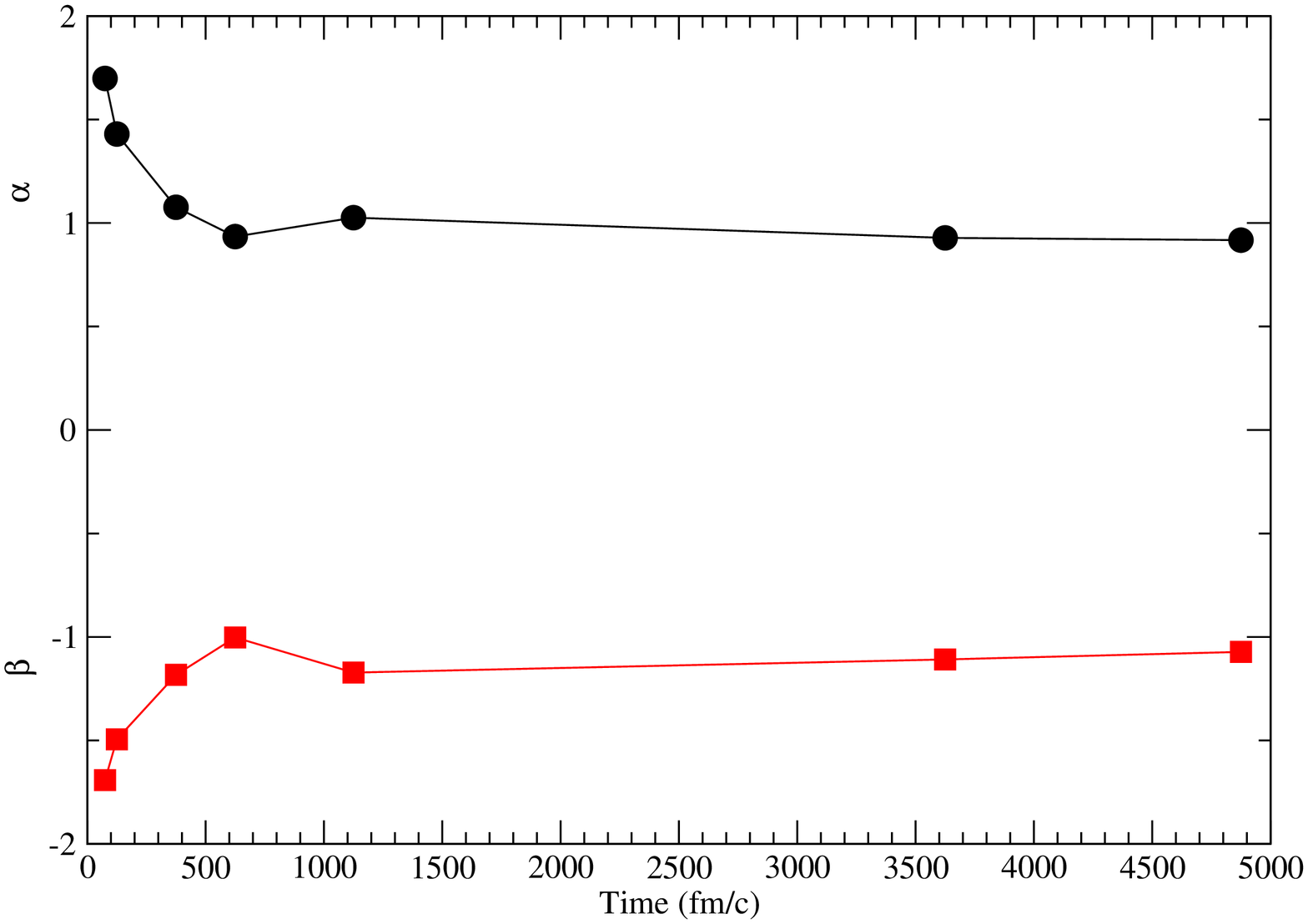}}
\caption{\label{fig4}Time evolution of $\alpha$ and $\beta$ for the
case ${}^{40}$Ca - ${}^{48}$Ca at $35 \ MeV/A$.}
\end{figure}

Repeating the fitting procedure described in~\ref{isoscaling}, it is
possible to obtain values of $\alpha$ and $\beta$ at different
times. Figure~\ref{fig4} shows the time evolution of $\alpha$ and
$\beta$ obtained from the ratio of yields of ${}^{40}$Ca -
${}^{48}$Ca at $35 \ MeV/A$ as a function of reaction time.

As attested by figure~\ref{fig3} and with varying values of $\alpha$
and $\beta$ ({\it cf.} figure~\ref{fig4}), isoscaling appears to be
present in the reaction from very early times. This, in agreement
with the findings of~\cite{tsang}, rules out long-time effects, such
as secondary decays, as the cause of isoscaling.  Thus, our search
for the origin of this effect should now be directed to earlier
times of the reaction, those in which the system is still very dense
and highly interacting.

\begin{figure}[htb]
\par
\centerline{ \includegraphics[scale=0.5,angle=-90]{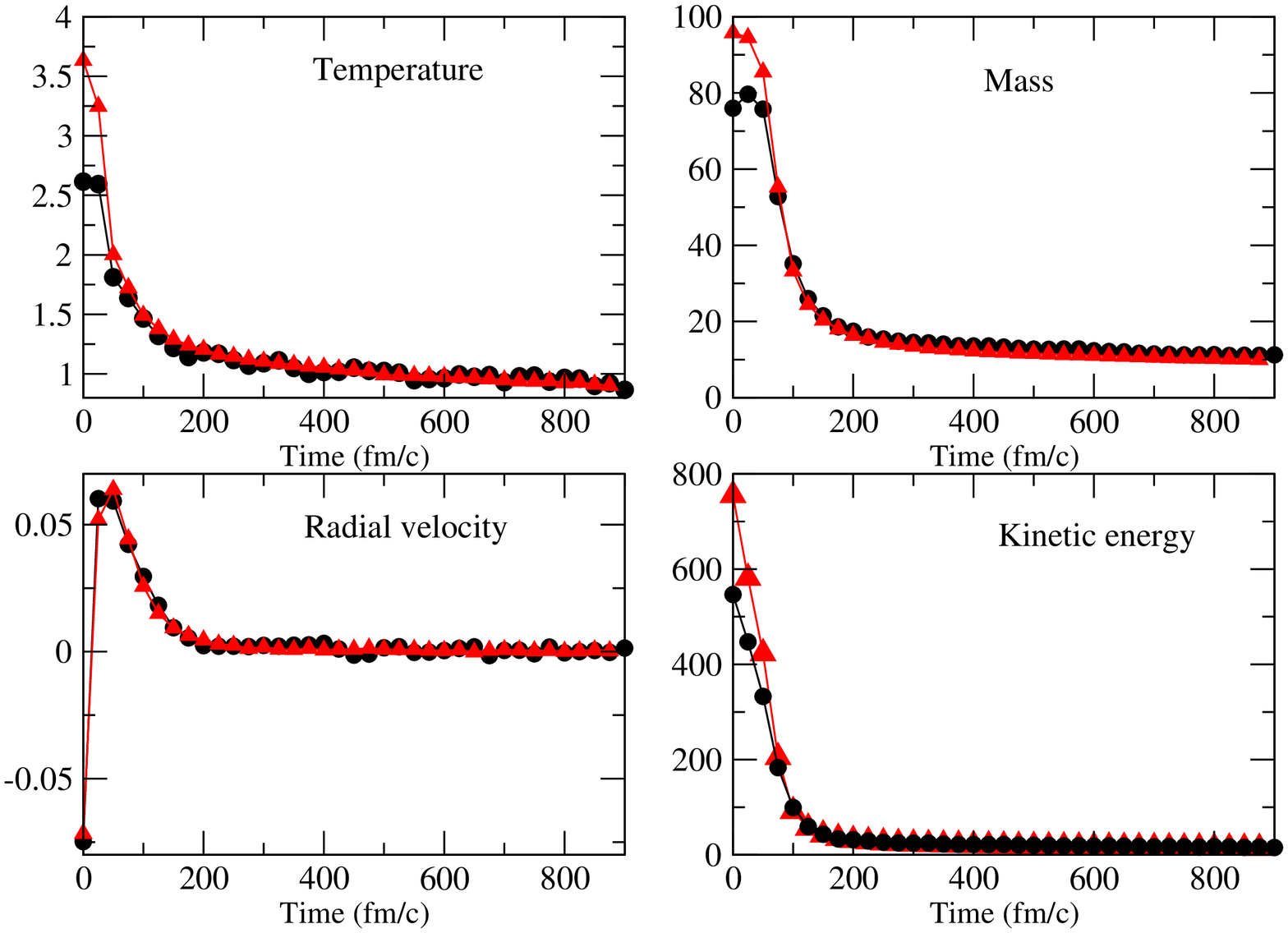}}
\caption{\label{fig5}Average properties of the biggest fragment.
(a) temperature, (b) mass, (c) collective radial velocity, and (d)
total kinetic energy. The circles correspond to collisions of
${}^{40}$Ca on ${}^{40}$Ca, and the triangles to ${}^{48}$Ca on
${}^{48}$Ca, both at $35 \ MeV/A$.}
\end{figure}

\subsection{Isoscaling and dynamics}\label{dynamics}
As outlined in reference~\cite{xu}, the parameters $\alpha$ and
$\beta$ of the power law~(\ref{r12}), can be linked to the
differences between the neutron and proton separation energies for
the two reactions. Assuming that the reactions populate a grand
canonical ensemble, and that the secondary decays have little
impact on the resulting $R_{21}$, it can be shown that
$\alpha=(\mu_{n2}-\mu_{n1})/T$ and $\beta=(\mu_{p2}-\mu_{p1})/T$,
where $\mu_{ni}$ and $\mu_{pi}$ are the neutron and proton
chemical potentials of reaction $i$, and $T$ is the equilibrium
temperature of the reaction, assumed to be the same for both
reactions of the isoscaling comparison.

The main assumption of the preceding arguments, namely, the
existence of thermodynamic equilibrium, is questionable in systems
which are finite, expanding and disassembling. Added to this
concern, of courses, is the assumption of a common temperature in
both reactions, as well as unique separation energies throughout the
disassembling systems.  Although these assumptions are difficult to
verify, next we extract related information to better understand
these premises.

As the temperature of a fragmenting system is not well defined, we
focus only on the dynamical properties of the largest fragment.
Figure~\ref{fig5} shows the time evolution of the average
temperature, mass, collective radial velocity, and total kinetic
energy of the biggest fragments obtained in one thousand collisions
of ${}^{40}$Ca on ${}^{40}$Ca (circles) and of ${}^{48}$Ca on
${}^{48}$Ca (triangles) both at $35 \ MeV/A$. The temperature was
calculated as two thirds of the {\it internal} kinetic energy, {\it
i.e.} of the amount of kinetic energy left after removing the energy
stored in the collective expansion. The maximum temperatures
observed (between $2.5$ and $3.5 \ MeV$) are consistent with those
derived from alternative analyses~\cite{gelbke}.  Analyses of other
reactions at different energies also yielded similar results.

Figure~\ref{fig5} indicates that the time evolution of the
dynamical observables of both reactions is, on average, very
similar except at very early times. In the beginning of the
reaction, while isoscaling sets in, the largest fragments of both
reactions suffer similar compressions and expansions, but with
very different average temperatures. As indicated by the rapid
increase and decrease of the mass and radial velocity, the largest
fragments of both types of reactions appear to undergo a fast
period of compression and expansion in the beginning of the
reaction.  But, although both systems appear to be in synchrony in
this compression-expansion, their average temperatures differ by a
factor of the order of $30 \%$.

Even though this difference in temperature vanishes for later times,
its importance cannot be underestimated as it occurs precisely when
the $n$ to $p$ ratios and, thus, $R_{21}$ are being established.
This discrepancy in temperatures casts serious doubts about the
proposed relationship between $\alpha$ and $\beta$ and the neutron
and proton chemical potentials. This would, indeed, lead to
relationships of the type $\alpha=(\mu_{n2}/T_2-\mu_{n1}/T_1)$ and
$\beta=(\mu_{p2}/T_2-\mu_{p1}/T_1)$.

Although the existence of different source temperatures could
possibly be handled~\cite{tsang2002}, the dependence of $\mu$ on $T$
($\mu (T)-\mu(0)\propto T^2$~\cite{lopdor}) combined with the rapid
cooling in early times, makes it difficult to argue for the
existence of well defined separation energies that could apply
throughout the reaction. At best, what exists at, say, $125 \ fm/c$
is the sum of fragments emitted at earlier times which might have
undergone subsequent decay. The final yield ratio $R_{21}$ can only
be said to be a product of the time integral of decays that occurred
under different values of the chemical potentials.

\begin{figure}[htb]
\par
\centerline{ \includegraphics[scale=0.5,angle=-90]{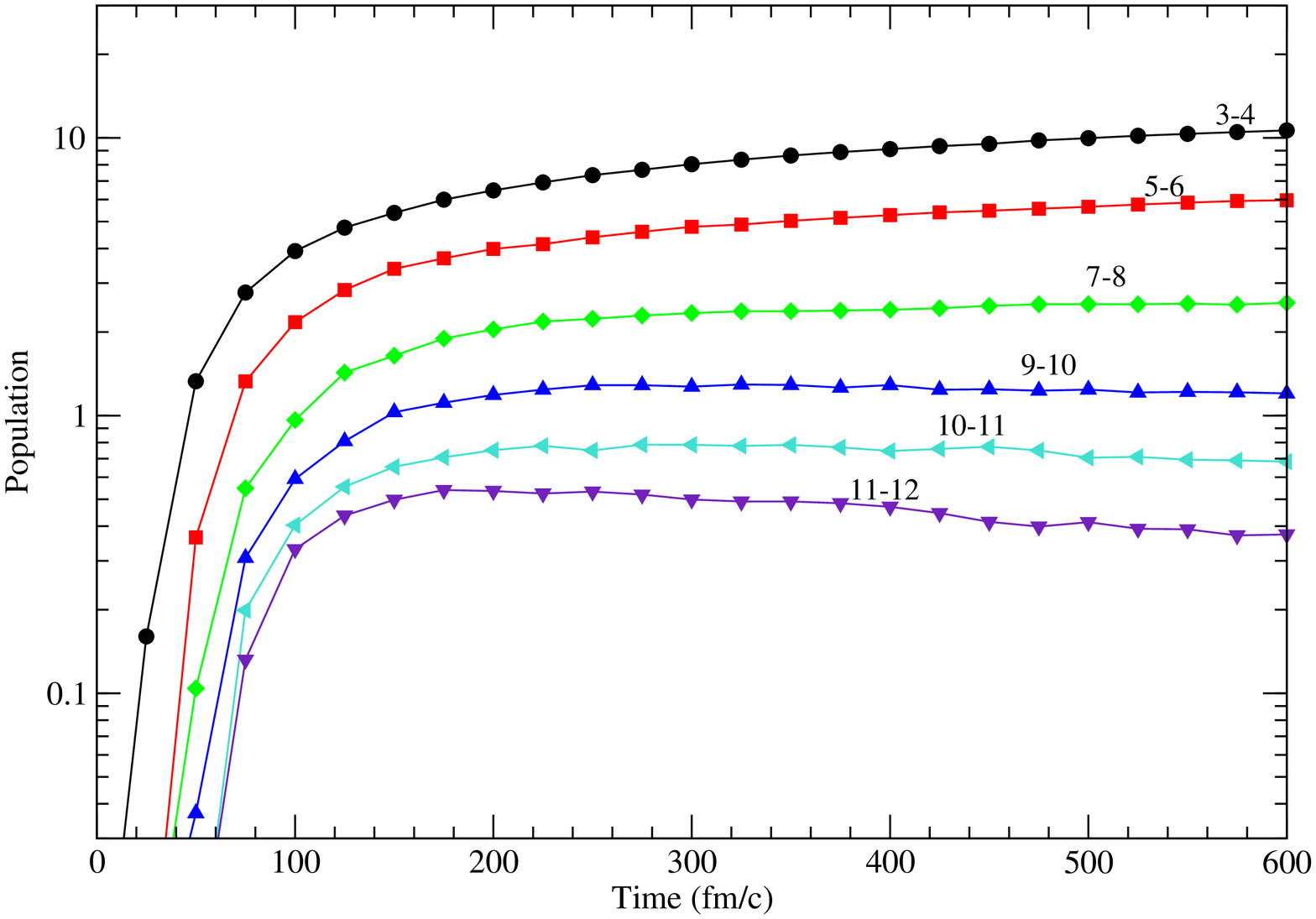}}
\caption{\label{fig6}Population of different mass bins as a
function of time for the reaction ${}^{40}$Ca on ${}^{40}$Ca at
$35 \ MeV/A$. Circles (3-4), squares (5-6), etc.}
\end{figure}

Besides ruling out the chemical potential as the decisive factor
in establishing isoscaling, the previous results introduce the
need to consider the addition of decays and other
nucleon-rearranging processes that take place at different times
during the reaction. The importance of this factor can be
quantified by inspecting the time evolution of the particle
production for different types of particles.

Figure~\ref{fig6} shows the average population of different mass
bins obtained at different times for the reaction ${}^{40}$Ca on
${}^{40}$Ca at $35 \ MeV/A$.  This time chart clearly illustrates
that at times smaller than $100 \ fm/c$, the fragment mass
distribution is far from reaching its final value. Furthermore,
since, as seen in figure~\ref{fig3}, isoscaling is already alive and
well at $125 \ fm/c$, {\it i.e.} well before the stabilization of
most mass bins, the isoscaling power law observed at early times is
only a work in progress produced by fragments emitted in previous
times.

By induction, all values of $\alpha$ and $\beta$ at a given time
include contributions of all earlier fragments, and will pass their
own contribution to later values of these exponents. It is, thus,
quite obvious that the isoscaling behavior is already present in the
initial -and strongly {\it out of equilibrium}- stages of the
evolution.  It does appear as if the origin of isospin is connected
more to the statistical sampling induced by the collision, as
observed by Ono {\it et al.}~\cite{ono}, than to specific details of
the dynamics of the reaction; this will be explored further in a
follow-up investigation~\cite{davila}.

\section{Conclusions}
The origin and evolution of isoscaling was studied using classical
molecular dynamics simulations combined with a fragment-recognition
algorithm. Collisions of ${}^{40}$Ca + ${}^{40}$Ca, ${}^{48}$Ca +
${}^{48}$Ca, and ${}^{52}$Ca + ${}^{52}$Ca, at energies from $20$ to
$85$ $MeV/A$ were used to construct the ratios $R_{21}(N,Z)$ and to
obtain the fitting parameters $\alpha$ and $\beta$ for the three
possible combinations of reactions at all energies and for times
from impact to $5000 \ fm/c$.

Isoscaling was found at asymptotic times from $1250 \ fm/c$ to $5000
\ fm/c$, the last calculated time. The fitting parameters of the
asymptotic $R_{21}$ showed a smooth variation with respect to the
beam energy. Although excellent agreement was found with quantal
molecular dynamics results, these classical results indicate that
the origin of isoscaling is not a quantum effect.

Isoscaling was detected at very early times during the collision
($50 \ fm/c$) and it was maintained, with smoothly varying values of
$\alpha$ and $\beta$, throughout the reaction. Examining the time
evolution of the fragment mass distribution, it is clear that
isoscaling exists well before this distribution reaches its final
value; whatever produces isoscaling is present in the very early
stage of the reaction.

Examining this period of time it was found that the largest
fragments of both reactions undergo a compression/heating phase
followed by an expansion/cooling period with average temperatures
which differ by up to $30 \%$. As this density-$T$ evolution implies
a strong variation of the separation energies, the relationship
between $\alpha$ and $\beta$ and a unique $\mu$, postulated in other
investigations, is obscured.

In summary, we found that isoscaling exists in classical systems, it
can be produced in dense systems out of equilibrium, it is
accumulative in time and it is maintained in systems expanding and
fragmenting. Thus, the existence of isoscaling appears to be
independent of the specific characteristics of the dynamics of the
reaction. Isoscaling must owe its existence to factors left
unexplored in this study, namely, purely geometrical and sampling
factors. This will be considered in our following
study~\cite{davila}.

\begin{acknowledgments}
C.O.D. acknowledges the support of a grant from the Universidad de
Buenos Aires, CONICET through grant 2041, and the hospitality of the
University of Texas at El Paso where this project was initiated. The
authors are indebted to A. Barra\~n\'on for facilitating the initial
configurations of the ``{\it nuclei}'' used in these simulations.
\end{acknowledgments}

\end{document}